\documentclass[aps,prd,showpacs,amssymb,superscriptaddress, notitlepage,floats,floatfix,nofootinbib, onecolumn]{revtex4-2}

\usepackage{graphicx}
\usepackage{dcolumn}
\usepackage{bm}
\usepackage{amsmath}
\usepackage{color}
\usepackage{hyperref}
\usepackage{subfigure}
\usepackage{lineno,hyperref}
\usepackage{nicematrix}
\modulolinenumbers[5]
\begin{document}
\title{Three-partite vertex model and knot invariants}

\author{T.K. Kassenova}
\email{tkkasenova@gmail.com}
\affiliation{L.N. Gumilyov Eurasian National University, Nur-Sultan, Kazakhstan}
\affiliation{M.Kh.Dulaty Taraz Regional University, Taraz, Kazakhstan}

\author{P. Tsyba}
\email{pyotrtsyba@gmail.com}
\author{O. Razina}
\email{olvikraz@gmail.com}
 
\affiliation{L.N. Gumilyov Eurasian National University, Nur-Sultan, Kazakhstan}

\author{R. Myrzakulov}
\email{rmyrzakulov@gmail.com}
\affiliation{L.N. Gumilyov Eurasian National University, Nur-Sultan, Kazakhstan}
\affiliation{LLP Ratbay Myrzakulov Eurasian International Center for Theoretical Physics, Nur-Sultan, Kazakhstan}

\date{\today}

\date{\today}

\begin{abstract}
This work dedicated to the consideration of the construction of a representation of braid group generators from vertex models with $N$-states, which provides a great way to study the knot invariant. An algebraic formula is proposed for the knot invariant when different spins $(N-1)/2$ are located on all components of the knot. The work summarizes procedure outputting braid generator representations from three-partite vertex model. This representation made it possible to study the invariant of a knot with multi-colored links, where the components of the knot have different spins. The formula for the invariant of knot with a multi-colored link is studied from the point of view of the braid generators obtained from the $R$-matrices of three-partite vertex models. The resulting knot invariant $5_2$ corresponds to the Jones polynomial and HOMFLY-PT.
\end{abstract}

\maketitle

\section{Introduction}

		An impotant event wich happened at the end of the 20th century in the mathematical theory of knots and links, is the discovery of new and direct ways of constructing knot invariants using the statistical mechanics method. Since knots and links can be received from the closure of the braid word, knot and link invariants may be obtained from the representative theory of braid groups. There are many ways to solve the representation of different braid groups.
By Drinfeld's quantum group, as well as the independent research of F.A. Berezin \cite{Berezin2008198} in the field of statistical and quantum mechanics, the statistical sum turned out to be closely related to the invariants of knots (and links).
Vertex and IRF models not only allow us to very easily reconstruct the Jones polynomial from the statistical sum, but actually lead to a whole new series of Jones-type knot invariants \cite{Murasugi1996249}.
By the help of the Ising model in the theory of phase transitions and Hill's theory in the theory of liquids \cite{Gaudin19838} and liquid solutions, significant advances have been made in numerous application areas of statistical mechanics \cite{Hill19879}. Consideration of the tetrahedron equation or the Zamolodchikov equation, which is a 3D generalization of the well-known Yang-Baxter equation \cite{Stroganov1997141}, shows that it provides invariance with respect to Reidemeister III motions, the importance of charge conservation and the construction of knot invariants from vertex models \cite{YangGe1989452}. Heretofore, statistical mechanics and knot theory had nothing to do with each other. However, in the process of discovering new polynomial invariants for knots, Vaughan Jones established a connection between these two fields. Currently, this is of research \cite{Adams2004205} is extremely active \cite{EkholmLenhardSullivan2013975}. But the question remains open whether there are Vassiliev invariants that can distinguish an oriented knot from its inverse, i.e., a knot with an opposite orientation \cite{Sawollek2003767}.
Families of polynomial invariants virtual knots and links arise when considering some $2\times2$ matrices with elements quaternions \cite{BartholomewFenn2008231}.

In the next paper, the Jones polynomial $VL(t)$ is calculated for several families of alternating knots and links, and the Tutte polynomial $T(G,x,y)$ for the associated graphs $G$, and the computation of the Jones polynomial for unchanging links is also discussed \cite{ChangShrock2001196}.

The research relationship between knots and braids provides the process of obtaining a knot or link from a given braid by “closing” the braid, which leads to the formulation of two fundamental questions about knots and braids \cite{BirmanBrendle20041}.
The study of welded braids and classical knots \cite{Kamada2007441} and a new invariant of virtual knots and links \cite{Kauffman20041} have led to the rapid development of knot theory. 

The theory of module skeins is a consequence of the discovery the Jones polynomial \cite {KauffmanLambropoulouJablan2001407}, which led to a thorough analysis of all other classes of $2$-bridge knots \cite{SiwachMadeti20151}. Creation an invariant of virtual knots and links \cite{DyeKauffman20091335} and the definition of the properties of birac and biquandl, where natural invariants arise from virtual knots and braids \cite{FennJordan-SantanKauffmana2004157}.
The new virtual knot invariant and the transcendental functional invariant combine several polynomial virtual knot invariants, such as the zero polynomial \cite{Jeong20151550078,Cheng20151}, the bend of the polynomial \cite{ChengGao20131341002}, and the affine index polynomial \cite{Kauffman20131341002}. 

It also explains the general method for calculating the Tutte polynomial for partitioning a graph \cite{JinZhang2003391}. The article computes exact expressions for the Jones polynomials for a family of links \cite{JinZhang2004183}.

According to the Chern-Simons $SU(2)$ theory, Witten's pioneering work \cite{Witten1989351} is one such method in which the Jones polynomial located at the knot $K$ \cite{Jones1997448} is formed from the expectation of the Wilson loop $\left\langle W_R(K)\right\rangle$.
A great quality of this approach is the relationship between the 3D $SU(2)$ Chern-Simons  theory and the two-dimensional $SU(2)_k$ Wess-Zumino-Witten (WZW) model, where $k$ is the Chern-Simons coupling constant showing degree of the WZW model. For example, the deformation coefficient $q=exp(2\Pi/k + 2)$ for the group $SU(2)$ \cite{KirillowReshetikhin1989285}.
The Temperley-Lieb algebra and the Birman-Murakami-Wenzl algebra (BMW) played a main role in solving models and knot theory \cite{Gepner2020115116}.

The article mentions especially of the most important and well-known classical results of F.Wu, and also describes some of his recent research on the connection of lattice statistical mechanical models with deep problems of pure mathematics  \cite{Maillard200328}.

The description of the Boltzmann weight and the determination of $SO(N)$ for any $N$ spin vertex model of algebra \cite{BelavinGepnerWenzl2020115160} served the discovery of new problems in the field of statistical physics. Braid group representations obtained from rational conformal field theories can be used to obtain explicit representations of Temperley-Lieb-Jones algebras \cite{Kaul1994267}. It was found that the Yang-Baxter relation provides both an algebraic and a graphical method in knot theory \cite{DeguchiWadatiAkutsu1989193}.
The universal $R$-matrices constructed using generators $J^\pm, J_z$ of matrices $U_q(SU(2))$ obey the defining relations of braid groups.
Therefore, these universal $R$-matrices are representations of braid groups.
Exactly solvable models of statistical mechanics \cite{Baxter19823} features from the method of constructing a representation of the braid group. Vertex models of $N$-states are one of such statistical mechanical models with Boltzmann weights $(R^{j,j})^{n_1,n_2}_{m_1,m_2}(u)$ associated with each vertex on a square lattice, depending on the spins $m_1, m_2, n_1, n_2\in\left\{-j,-j+1,\ldots j\right\}$, located on four edges intersecting the vertex, where spin $j=(N-1)/2$. The Yang-Baxter equations obey the spectral parameters $u$ depending on the Boltzmann weight $(R^{j,j})^{n_1, n_2}_{m_1, m_2}(u)$ of these vertex models.
 In fact, the Yang-Baxter equation in the limit $u\rightarrow\infty$ can be reduced to the definition of braid group relations by applying the permutation operator $\hat{P}$ to the Boltzmann weights.
This means that, the braid generators are proportional to $\hat{P}[(R^{j,j})^{n_1, n_2}_ {m_1, m_2}(u\rightarrow\infty)]$. The Boltzmann weights for new vertex models in which edges carry states with spin $j>3/2$ \cite{Kauffman20131341002} ensure us with new representations of braid matrices that are productive for constructing new knot invariants using the algebraic expression $\alpha_{\underbrace{j, j, \ldots j}_{n}}(A)$ \cite{AkutsuWadati1988243,AkutsuWadati19873039,AkutsuDeguchiWadati19873464,DeguchiAkutsuWadati1988757,Deguchi19893441,DeguchiWadatiAkutsu19882921}.
Application of algebraic expression when obtaining the polynomials of the knot and links is the optimal method corresponding to any arbitrary word braid $A$ and includes only the multiplication of matrices. Traditional approach vertex model is able to efficiently compute the knot and link invariant $\alpha_{\underbrace{j, j, \ldots j}_{n}}(A)$. An attempt to obtain a modified algebraic expression for the knot invariant  $\alpha_{\underbrace{j, j, \ldots j}_{n}}(A)$ in terms of matrix representations of braid matrices were called multi-colored knot and link invariants, where $j_1, j_2,\ldots$-spin states located in different constituent links of the knot. The mathematical solution for the knot polynomial using these braid matrices was studied in publications \cite{AkutsuWadati1988243,AkutsuWadati19873039,AkutsuDeguchiWadati19873464,DeguchiAkutsuWadati1988757,Deguchi19893441,DeguchiWadatiAkutsu19882921} for the spin $j=1/2, j=1, j=3/2$, which are common in the sources as the $6$-vertex, $19$-vertex and $44$-vertex models. The numbers $6,19,44$ indicate the number of nonzero Boltzmann weights for the corresponding vertex model with $N$-states.

	As a result of obtaining an eight-vertex model using the Grassmann algebra, a series of solutions to the tetrahedron equation is specified as a sufficient condition for the permutation of the transition matrices on a simple square lattice \cite{KassenovaTsybaRazina2019012035}. In the fifteen-vertex model, which satisfying the ice rule, the algebraic-differential method \cite{Vieira201919} is utilized to solve the Yang-Baxter equation, where the matrix elements also depend on the spectral parameter.
The $32$-vertex free fermion model, which demonstrates the Ising behavior, can also has various transitions \cite{SaccoWu1975}. 

The article examines a family of solvable A-D-E lattice models that demonstrate order-disorder transitions, as well as different classes of critical behavior universality \cite{Pearce199415}.

The Ashkin-Teller model represents two Ising models ($s$ and $\sigma$)-models. Subsequently, the Ashkin-Teller model on a square lattice attracted the attention of physicists. Later it turned out that it is conformally invariant. The close relationship of the Ashkin-Teller model on a square lattice was studied with models of statistical mechanics such as like the Potts model, the eight-vertex model with two sublattices is a special case of the six-vertex model \cite{Gusev1990396}. Using the method of diagonalization by known Boltzmann weights for the $15$-vertex model, $32$-vertex model, and the Ashkin-Teller model on a square lattice, one can obtain diagonal matrix elements $\lambda_J (j_1, j_2, j_3; u)$, depending on the spectral parameter. In vertex models with Boltzmann weights, when diagonalizing, one can verify that the diagonal matrix elements depend on the spectral parameter for the spin $j>3/2$ \cite{Kaul1998}. 

		The method for obtaining invariants is associated with multi-colored links \cite{KaulGovindarajan1993392}, with the statistical sum of the Chern-Simons theory \cite{KaulRamadevi2001295} and with obtaining the topological solution $SU(2)$ of the Chern-Simons theory on $S^3$ \cite{Kaul1994289}. The 3D $SU(2)$ Chern-Simons theory has been studied as a topological field theory for the field-theoretical description of knots and links in three dimensions \cite{Gusev1990396}. Thus, one can use braid theory to study knot theory and vice versa \cite{Kamada200211}.
		Thereafter, the connection between the invariant of knot theory and the new ten-vertex model of statistical mechanics was studied, with using the transition of the commuting transfer matrix, including the Boltzmann weights, into the braid matrix \cite{KassenovaTsybaRazina2020694}.

The polynomial form can be calculated for any braid word $A$ using the algebraic expression $\alpha_{\underbrace{j, j, \ldots j}_{n}}(A)$, which will be considered below in the formula $(1.4)$, as described in \cite{AkutsuWadati1988243,AkutsuWadati19873039,AkutsuDeguchiWadati19873464,DeguchiAkutsuWadati1988757,Deguchi19893441,DeguchiWadatiAkutsu19882921}. Here, the spin states $j$ are located on all strands of the braid, and $n$ denotes the number of knots components of the link obtained as a result of the closure of the braid $A$. We need to map these braid matrices to the monodromy matrices in the WZW models.
Braid generators should be derived from $(R^{j_1, j_2, j_3})^{n_1, n_2, n_3}_{m_1, m_2, m_3}(u)$-matrices of new vertex models. We call such vertex models three-partite vertex models in which each strand has $ m_1, n_1 \in j_1, j_1-1, \ldots-j_1 $, $ m_2, n_2 \in j_2, j_2-1, \ldots-j_2 $, $ m_3, n_3 \in j_3, j_3-1, \ldots-j_3 $. The procedure for obtaining Boltzmann weights for the vertex model $(R^{j_1, j_2, j_3})^{n_1, n_2, n_3}_{m_1, m_2, m_3}(u)$ from the spectral parameter-dependent elements of the diagonal braid $\lambda_J(j_1, j_2, j_3)$ can be generalized to obtain three-partite vertex models of Boltzmann weights $(R^{j_1, j_2, j_3})^{n_1, n_2, n_3}_{m_1, m_2, m_3}(u)$, where each strand carries different spins $j_1, j_2, j_3$. It is important to emphasize that the relations of quasigroups \cite{Witten1989351} for the braid generators $\hat {P}[(R^{j_1, j_2, j_3})^{n_1, n_2, n_3} _ {m_1, m_2, n_3}(u\rightarrow\infty)]$ were obtained from the Boltzmann weights in the limit $u\rightarrow \infty$ using the permutation operator $\hat{P}$.

		Our main aim in this article is to construct braid representations from $(R^{j_1, j_2, j_3})^{n_1, n_2, n_3}_{m_1, m_2, m_3}(u)$, which obey the properties of quasigroups \cite{Kaul1994289} from various spin $j$-states on the strands. This paper describes the derivation of the matrix form of the three-partite vertex model from the Boltzmann weights, which allows using the modified formula for the knot from the closure of an arbitrary word of the braid $A$ and effectively calculate the multi-colored knot invariant and link $\alpha_{\underbrace{j, j,\ldots j}_{n}}(A)$.
This work is dedicated to graphical methods of calculation, which are based on the Clebsch-Gordan coefficients and transformation matrices \cite{JucysBandzaitis1965531}. Using the quantum-deformed Clebsch-Gordan coefficients $q-CG$, these braid matrices $\hat{P}[(R^{j, j})^{n_1, n_2}_{m_1, m_2}(u\rightarrow\infty)]$ can be diagonalized \cite{KaulGovindarajan1992293}, whose diagonal elements $\lambda_J(j, j)$ are the eigenvalues of the monodromy matrices in the WZW model. 

		The work plan is as following. In Section 2, we consider the construction of braid matrices from the $R$-matrix of the usual and ten-vertex models for the same case of spin and the derivation of the knot invariant with links.
In Section 3, we generalize the procedure for defining a new representation of the braid generators from the $R(u)$-matrix associated with the three-partite vertex model and propose an algebraic formula for the invariant of a knot with multi-colored links.

\section{Vertex model \& R-matrix}

In this section, we briefly review the ten-vertex model approach to constructing representations of braid group generators leading to the computation of the $b_1^5$ knot invariant.

\paragraph{Vertex model}

$N$-vertex models consider a two-dimensional model of statistical mechanics with states of the same spin $j$, located on four edges intersecting each vertex, as shown in figure $1$.
\begin{figure}
\centering
\includegraphics[scale=1]{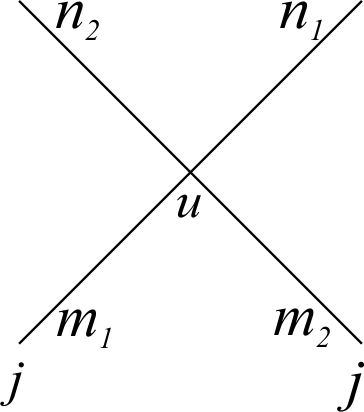}
\caption{Vertex model $(R^{j,j})_{m_1,m_2}^{n_1,n_2}(u)$}
\end{figure}

This exactly solvable model of statistical mechanics shows that the spectral parameter $u$ directly depends on the Boltzmann weights $(R^{j,j})_{m_1,m_2}^{n_1,n_2}(u)$ and with the Yang-Baxter equation is agreed  
\begin{eqnarray}
\Sigma_{m_{1}^{'},m_{2}^{'},m_{3}^{'}}(R^{j,j})_{m_1,m_2}^{m_1^{'},m_2^{'}}(u)(R^{j,j})_{{m_1^{'},m_3}}^{{m_1^{''}},{m_3^{'}}}(u+\upsilon)(R^{j,j})_{m_2^{'},m_3^{'}}^{m_2^{''},m_3^{''}}(\upsilon)=\nonumber\\=\Sigma_{m_{1}^{'},m_{2}^{'},m_{3}^{'}}(R^{j,j})_{m_2,m_3}^{m_2^{'},m_3^{'}}(\upsilon)(R^{j,j})_{{m_1,m_3^{'}}}^{{m_1^{'}},{m_3^{''}}}(u+\upsilon)(R^{j,j})_{{m_1^{'}},m_2^{'}}^{m_1^{''},m_2^{''}}(u).
\end{eqnarray}
The parameterized form of this $R$-matrix depends on the spectral parameter $ u $ and the deformation parameter $q=e^{2\mu}$ is given in \cite{Baxter19823,AkutsuWadati1988243} for $6,9,44$-vertex models. The above equation in the limit $u\rightarrow 0$, supplementing the product of $R$-matrix elements by the permutation operator $\hat{P}$ (up to full normalization), shows an explicit relation for the braid group $B_r$, where the generators $b_i$'s $(i=1,2,\ldots r)$ have the following representation
\begin{eqnarray}\label{Knot2}
b_{i}[j,j]=\underbrace{\mathbb{I}_{1}\times\mathbb{I}_{2}\times\ldots\mathbb{I}\times}_
{i-1}(\hat{R}^{j,j})_{m_1,m_2}^{n_1,n_2}\times\mathbb{I}_{i+2}\ldots,
\nonumber\\
b_{i}[j,j]^{-1}=\underbrace{\mathbb{I}_{1}\times\mathbb{I}_{2}\times\ldots \mathbb{I}\times}_{i-1}((\hat{R}^{j,j})_{m_1,m_2}^{n_1,n_2})^{-1}\times \mathbb{I}_{i+2}\ldots. 
\end{eqnarray}
Here 
\begin{eqnarray}\label{Knot3}
((\hat{R}^{j,j})_{m_1,m_2}^{n_1,n_2})=\frac{1}{N}\hat{P}((R^{j,j})_{m_1,m_2}^{n_1,n_2})(u\rightarrow\infty),
\end{eqnarray}
normalization factor $N=(R^{j,j})_{j,j}^{j,j}(u\rightarrow\infty)$ ensures that all matrix elements in the limit $u\rightarrow\infty$. As a result, we have a new representation of the braid of the $R$-matrix of the vertex model, which leads to new invariants of the knot or link. The mathematical form $\alpha_{\underbrace{j,j,\ldots j}_{n}}(A)$ \cite{AkutsuWadati1988243} helps to calculate the invariant of any $n$-component link at a knot obtained by the closure of the braid word $A\in B_r$, with the same spin $j$ at other multicomponent sites 
\begin{eqnarray}\label{Knot4}
\alpha_{\underbrace{j,j,\ldots j}_{n}}(A)=(\tau_{j}\overline{\tau}_j)^{-n/2}(\frac{\overline{\tau}_j}{\tau_j})^{e/2}Tr[HA],
\end{eqnarray}
where $e$ is the exponent sum of the $b_i$'s appearing in the braid word $A$, \\
$H=\underbrace{h_j\otimes h_j\ldots h_j}_{r}$
where
\begin{eqnarray}\label{Knot5}
h_j=\frac{1}{1+q+\ldots+q^{2j}}Diag[1,q,\ldots,q^{2j}],
\end{eqnarray}
and $\tau_j$ and $\overline{\tau}_j$ are
\begin{eqnarray}\label{Knot6}
\tau_j=\frac{1}{1+q+\ldots+q^{2j}};\ \ \overline{\tau}_j=\frac{q^{2j}}{1+q+\ldots+q^{2j}}.
\end{eqnarray}
When performing Markov movements on the braid, the above invariant $\alpha_{j,j,\ldots j}(A)$ with the variable $q$ remains unchanged, and for a trivial knot, the braid word $A$ has the form $\alpha_{j}(b_1)=\sum_{i=-j}^{j}q^{i}$.
In sources on knot theory, this invariant is known as unnormalized knot invariants.

  In the next subsection, we briefly study the computation of the knot invariant using the braid matrices obtained from the $R$-matrices of the new model with $10$ vertices. It is assumed here that the edges in fig. $1$ have spins $j=1/2$.
 
\paragraph{Ten-vertex model} 

The new vertex model is a ten vertex model in which the spins $j=1/2$ are located on four edges that intersect each vertex. Thus, the Boltzmann weights $(R^{j,j})_{m_1,m_2}^{n_1,n_2}(u)$ associated with each vertex are nonzero if and only if $m_1+m_2=n_1+n_2$ where $m_1, m_2,n_1,n_2\in ({-1/2, 1/2})$. This condition admits ten nonzero Boltzmann weights, resulting in a model with $10$ vertices. In matrix form, the elements are
\begin{eqnarray}
\resizebox{.8\textwidth}{!}{$R_{m_1,m_2}^{n_1,n_2}=\begin{pmatrix}R_{m_1,m_2}^{n_1,n_2}&\vline&\uparrow\uparrow&\downarrow\uparrow&\uparrow\downarrow&\uparrow\downarrow&\downarrow\downarrow\\
\hline\uparrow\uparrow
&\vline&\sinh(\lambda-u)&\sinh 2\lambda \sinh u&0&0&\sinh(\lambda)\sinh(u-\lambda)\\\downarrow\uparrow
&\vline&\sinh(\lambda+u)&\sinh\lambda&\sinh\lambda&0\\\uparrow\downarrow
&\vline&0&\sinh\lambda&\sinh(\lambda+u)&\sinh(\lambda+u)&0\\\uparrow\downarrow
&\vline&0&\sinh\lambda&\sinh(\lambda+u)&\sinh(\lambda+u)&0\\\downarrow\downarrow
&\vline&\sinh(\lambda)\sinh(u-\lambda)&0&0&\sinh 2\lambda \sinh u&\sinh(\lambda-u)
\end{pmatrix}$}.
\end{eqnarray}
To construct the braid generators $b_i$, we take the spectral parameter, which is considered for $u\rightarrow\infty$ by the above matrix elements and replace $e^{2\lambda}$ with the variable $q$. Compare with the corresponding normalization so that the matrix elements are finite in this limit $u\rightarrow\infty$, as shown below

\begin{eqnarray}
\resizebox{.8\textwidth}{!}{$\frac{(R^{\frac{1}{2},\frac{1}{2}})_{m_1,m_2}^{n_1,n_2}(u\rightarrow\infty)}{(R^{\frac{1}{2},\frac{1}{2}})^{\uparrow,\uparrow}_{\uparrow,\uparrow}(u\rightarrow\infty)}=
\begin{pmatrix}
R_{m_1,m_2}^{n_1,n_2}&\vline&\uparrow\uparrow&\downarrow\uparrow&\uparrow\downarrow&\uparrow\downarrow&\downarrow\downarrow\\
\hline\uparrow\uparrow
&\vline&1&\frac{1}{2}(-\sqrt{q+q^{\frac{3}{2}}})&0&0&\frac{-1+q}{2\sqrt{q}}\\\downarrow\uparrow
&\vline&0&0&\sqrt{q}&\sqrt{q}&0\\\uparrow\downarrow
&\vline&0&\sqrt{q}&-1+q&-1+q&0\\\uparrow\downarrow
&\vline&0&\sqrt{q}&-1+q&-1+q&0\\\downarrow\downarrow
&\vline&\frac{-1+q}{2\sqrt{q}}&0&0&\frac{1}{2}(-\sqrt{q+q^{\frac{3}{2}}})&1
\end{pmatrix}$}.
\end{eqnarray}
Applying the permutation matrix
\begin{eqnarray}
\hat{P}^{1/2, 1/2}=
\begin{pmatrix}
1&1&0&0&1\\
0&0&0&1&0\\
0&0&1&0&0\\
0&1&0&0&0\\
1&0&0&1&1
\end{pmatrix}
\end{eqnarray}
the elements of $(\hat{R}^{\frac{1}{2},\frac{1}{2}})_{m_1,m_2}^{n_1,n_2}$ \eqref{Knot3}turn out to be
\begin{eqnarray}
(\hat{R}^{\frac{1}{2},\frac{1}{2}})_{m_1,m_2}^{n_1,n_2}=
\begin{pmatrix}
1& \frac{1}{2}(-\sqrt{q+q^{\frac{3}{2}}})& 0& 0& \frac{-1+q}{2\sqrt{q}}\\
0& \sqrt{q}& 0& \sqrt{q}& 0\\
0& \sqrt{q}& -1+q& \sqrt{q}& 0\\
0& -1+q& -1+q& -1+q& 0\\
\frac{-1+q}{2\sqrt{q}}& 0& 0& \frac{1}{2}(-\sqrt{q+q^{\frac{3}{2}}})& 1
\end{pmatrix}.
\end{eqnarray}
Thus, it is possible to compute the matrix form of the braid generators $b_i [1/2, 1/2]$ \eqref{Knot2}, using the $\hat{R}$-matrix. Using the formula \eqref{Knot4}, one can calculate the invariant for some knot with links. Be aware that there is only one braid generator $b_1$ for all braid words $A\in B_2$, whose matrix form will be $5\times 5$. It,
\begin{eqnarray}
b_1=({R}^{\frac{1}{2},\frac{1}{2}})_{m_1,m_2}^{n_1,n_2}.
\end{eqnarray}
Equation \eqref{Knot4} shows a ready-made invariant for links of an infinite group of knots, trefoils and Hopf using the braid words $A=b_1,b_1^3$ and $b_1^2$ respectively \cite{Deguchi19893441}. For knots and links resulting from the closure of braid words $A\in B_3$, there are two braid generators $b_1, b_2$, which are $10\times 10$ matrices
\begin{eqnarray}
b_1={R}^{\frac{1}{2},\frac{1}{2}})_{m_1,m_2}^{n_1,n_2}\times\mathbb{I},
\end{eqnarray}
\begin{eqnarray}
b_2=\mathbb{I}\times{R}^{\frac{1}{2},\frac{1}{2}})_{m_1,m_2}^{n_1,n_2}. 
\end{eqnarray}
For example, the knot $6_1$, in which the word braid is $A=b_1b_2^{-1}b_3b_4^{2}b_3^{-1}b_4^{-1}$. You should know that such an action of the braid word on a $3$-strand braid implies the following order of work of the matrix on the initial state $\left|\right.j, m_1; j,m_2; j;m_3\rangle$
\begin{eqnarray}\label{Knot14}
A|3-\text{strand}\rangle\equiv b_2\left[b_{1}^{-1}\left\{b_{2}(b_{1}^{-1}|j, m_1; j, m_2; j, m_3\rangle)\right\}\right].
\end{eqnarray}
The generalized way of solving the braid $A\in B_n$ for any word leads to the calculation of polynomial invariants in the formula \eqref{Knot4}. This polynomial is similar to the Jones polynomial, although the method includes only matrix product. The method is very effective in obtaining polynomial invariants for any knot or link from a vertex model, the elements of the $R$-matrix of which are known.
In the sources on knot theory, the multi-colored Jones polynomials correspond to the placement of the higher spins $j\geq 1$ on the components of the knot. Interestingly, these polynomials for $j=1,3/2$ agree with the knot and link invariant $\alpha_{\underbrace{j,j,\ldots j}}(A)$ in egn. \eqref{Knot4}, where the matrix representation of the generators of the braids $b_i$'s is obtained from the Boltzmann weights of the $19$-vertex and $44$-vertex models. 
The Wess-Zumino conformal field theory implies a compact relation for the braid generators $b_i$'s obtained from the $({R}^{j,j})_{m_1,m_2}^{n_1,n_2}$-vertex model matrices and also from the eigenvalues $(\lambda)$ of the monodromy matrix in $SU(2)_k$ \cite{Yang19671312}
\begin{eqnarray}\label{Knot15}
\hat{R}^{j,j})_{m_1,m_2}^{n_1,n_2}=\frac{1}{\textit{N}}\hat{P}^{j,j}({R}^{j,j})_{m_1,m_2}^{n_1,n_2}(u\rightarrow\infty)=\nonumber\\=\frac{1}{\textit{N}}\hat{P}^{j,j}\sum_{J\in j\otimes j}\begin{pmatrix}j& j& J \\ m_2& m_1& M\end{pmatrix}\lambda_J (j,j)\begin{pmatrix}j& j& J\\n_1& n_2& M\end{pmatrix},
\end{eqnarray}
where $M=m_1+m_2=n_1+n_2$ and items in brackets $\begin{pmatrix}j& j& J\\m_2& m_1& M\end{pmatrix}$ denote the quantum form of the formula for the Clebsch-Gordon coefficient $(q-CG)$ \cite{KirillowReshetikhin1989285}. In addition, the summation $J\in j\otimes j$ belongs to the range $\left\{0, 1, \ldots 2j\right\}$.
The essence of the main problem is to obtain the eigenvalues of $\lambda_J \left(j,j; u\right)$ depending on the spectral parameter, for any spin $j$ the relation in eqn. \eqref{Knot15} gives the value of $R^{j,j}(u)$-matrices, where matrix elements with spins $j_1=1/2, j_2=1,j_3=1/2$ belong to $15$-vertex model, $32$-vertex model and Ashkin-Teller models on a square lattice. This assumption $\lambda_J\left(j,j; u\right)$ is formulated in \cite{Yang19671312}
\begin{eqnarray}\label{Knot16}
\lambda_J(j,j;u)=\prod^{J}_{k_1=1}\left(\sinh(k_1\mu-u)\right)\prod^{2J}_{k_2=J+1}\left(\sinh(k_2\mu+u)\right),
\end{eqnarray}
as a result, matrices are obtained depending on the spectral parameter, where $(R^{j,j})_{m_1,m_2}^{n_1,n_2}(u)$ are matrices associated with new vertex models
\begin{eqnarray}\label{Knot17}
(R^{j,j})_{m_1,m_2}^{n_1,n_2}(u)=\sum_{J\in j\otimes j}\begin{pmatrix}j& j& J\\m_2& m_1& M\end{pmatrix}\lambda_J (j,j)\begin{pmatrix}j& j& J\\n_1& n_2& M\end{pmatrix}.
\end{eqnarray}
Here $SU(2)$ spin $J\in j\otimes j\equiv {0, 1, 2,\ldots 2j}$ (allowed irreducible representations in the tensor product).
We have verified for some spin values of $j$ that these $R$-matrices obtained from the assumed form in the formula \eqref{Knot16} obey the Yang-Baxter equation and hence are admissible Boltzmann weights for completely new vertex models. Initially, we analyzed the computation of the knot and link invariant from vertex models with edges carrying states with the same spin $j$. 

Therefore, one can see a specific relationship between the $R$-matrix with a spectral parameter that depends on the $\lambda_J(j,j;u)$ eigenvalue of the vertex model matrix. It is noteworthy that the assumed eigenvalue in the equation \eqref{Knot16} can be generalized to $\lambda_J(j_1,j_2,j_3;u)$, where $J\in j_1\otimes j_2\otimes j_3\equiv {|j_1-j_2-j_3|,|j_1-j_2-j_3|+1,\ldots j_1+j_2+j_3}$, which will lead to a vertex model with adjacent edges having states of different spins $j_1\neq j_2$, but $j_1=j_3$. These vertex models should be called three-partite vertex models. In the next section, we will describe the three-partite vertex model and propose a new algebraic expression for the multi-colored invariant of a knot with links from associated Boltzmann weights.

\section{Three-partite vertex model} 

In this paper, we investigate a new vertex model with different spins on adjacent edges of the lattice, which is called the “three-partite vertex model”. The eigenvalue in the formula \eqref{Knot16}  both for identical spins and for multi-colored spins $\lambda(j_1,j_2,j_3;u)$ has the form \cite{JucysBandzaitis1965531} 
\begin{eqnarray}\label{Knot18}
\lambda_J(j_1,j_2,j_3;u)=\prod^{J}_{k_1=|j_2-j_3|+1}\sinh\left(k_1\mu-u\right)\prod^{|j_1-j_3|}_{k_2=J+1}\sinh\left(k_2\mu-u\right)+\nonumber\\+\prod^{J}_{k_3=|j_2+j_3|+1}\sinh\left(k_3\mu+u\right)\prod^{|j_1-j_2|}_{k_3=J+1}\sinh\left(k_4\mu+u\right),
\end{eqnarray}
where $J\in j_1\times j_2\times j_3$ elements corresponding to the $R$-matrix depending on the spectral parameter (similar to the equation \eqref{Knot17})takes the form
\begin{eqnarray}\label{Knot19}
(R^{j_1,j_2,j_3})^{n_1,n_2,n_3}_{m_1,m_2,m_3}(u)=\nonumber\\=\sum_{j_{12},m_{12}}\begin{Bmatrix}j_2& j_1& J_{12}\\-m_2& -m_1& M_{12}\end{Bmatrix}\lambda_{J_{12}} (j_1,j_2;u)\begin{Bmatrix}j_1& j& J_{12}\\n_1& n_2& n_{12}\end{Bmatrix}+\nonumber\\+\sum_{j_{23},m_{23}}\begin{Bmatrix}J_{123}& j_1& j_{23}\\-m_{123}& m_1& m_{23}\end{Bmatrix}\lambda_{j_{23}}(j_1,J_{123};u)\begin{Bmatrix}j_1& J_{123}& j_{23}\\n_1& -n_{123}& n_{23}\end{Bmatrix}+\nonumber\\+\sum_{j_{23},m_{23}}\begin{Bmatrix}j_2& j_3& J_{23}\\m_{2}& m_3& -m_{23}\end{Bmatrix}\lambda_{J_{23}}(j_3,j_2;u)\begin{Bmatrix}j_3& j_2& J_{23}\\n_3& n_2& -n_{23}\end{Bmatrix}+\nonumber\\+\sum_{j_{12},m_{12}}\begin{Bmatrix}J_{123}& j_3& j_{12}\\m_{123}& -m_{3}& -m_{12}\end{Bmatrix}\lambda_{J_{12}} (J_{123},j_3;u)\begin{Bmatrix}j_3& J_{123}& j_{12}\\-n_3& n_{123}& -n_{12}\end{Bmatrix}.  
\end{eqnarray}
Earlier it was mentioned about the $R$-matrix satisfying the following Yang-Baxter equation \cite{Baxter19823,Yang19671312}
\begin{eqnarray}
\sum_{m^{'}_{1},m^{'}_{2},m^{'}_{3}}(R^{j_1,j_2,j_3})^{{m^{'}_{1},m^{'}_{2},m^{'}_{3}}}_{m_{1},m_{2},m_{3}}(u)(R^{j_1,j_3})^{{m^{''}_{1},m^{'}_{3}}}_{m_{1},m_{3}}(u+\upsilon)(R^{j_2,j_3})^{{m^{''}_{2},m^{''}_{3}}}_{m^{'}_{2},m^{'}_{3}}(\upsilon)=\nonumber\\=(R^{j_2,j_3})^{{m^{'}_{1},m^{''}_{3}}}_{m_{2},m_{3}}(\upsilon)(R^{j_1,j_3})^{{m^{'}_{1},m^{''}_{3}}}_{m_{1},m^{'}_{3}}(u+\upsilon)(R^{j_1,j_2})^{{m^{''}_{1},m^{''}_{2}}}_{m^{'}_{1},m^{'}_{2}}(u).
\end{eqnarray}
Investigation and calculation of the assumed form of the $R$-matrix \eqref{Knot19} for $j_1,j_2,j_3$ spins, completely satisfies the above Yang-Baxter equation. Starting from the limit $u,\upsilon,u+\upsilon\rightarrow\infty$ of $(R^{j_1,j_2})^{{m^{'}_{1},m^{'}_{2}}}_{m_{1},m_{2}}(u)$ and a suitable normalization $N=(R^{j_1,j_2,j_3})^{n_1,n_2,n_3}_{m_1,m_2,m_3}=(u\rightarrow\infty)$, we get matrix elements, not depending on the spectral parameter. Multiplying the corresponding permutation $\hat{P}^{j_1,j_2,j_3}$ by the $R$-matrix we obtain
\begin{eqnarray}
(\hat{R}^{j_1,j_2,j_3})_{m_1,m_2,m_3}^{n_1,n_2,n_3}=\frac{1}{\textit{N}}(\hat{P}^{j_1,j_2,j_3})^{{m^{'}_{1},m^{'}_{2},m^{'}_{3}}}_{{m_{1},m_{2},m_{3}}}({R}^{j_1,j_2,j_3})_{{m^{'}_{1},m^{'}_{2},m^{'}_{3}}}^{n_{1},n_{2},n_{3}}(u\rightarrow\infty),
\end{eqnarray}
calculating the braid generators $b(j_1,j_2,j_3)$, where the closure of each link in the knot occurs on three strands with representations $j_1,j_2,j_3$ 
\begin{eqnarray}
b(j_1,j_2,j_3)\left.|j_1,j_2,j_3\right\rangle\infty\left.|j_3,j_2,j_1\right\rangle.
\end{eqnarray}
An arbitrary braided word using these generators should track the spins of $j_1,j_2,\ldots j_n$ on $n$-strands. The collection of such braid words actually forms quasigroups \cite{JucysBandzaitis1965531}.

In addition, the closure requires that the initial state $\left|j_1,j_2,\ldots j_n\right\rangle$ remain the same as the final state after the braid word procedure. For example, closing a braid word will result in different knot and link components that carry different representations. Using the matrix form of the braid generators $b(j_1,j_2),b (j_1,j_3)\ldots$, obtained from three-partite vertex models, we reveal the invariant of multi-colored links for the components of the knot having different representations. In the next section, introducing explicit forms of Boltzmann weights, one can perform $R^{j_1=1/2, j_2=1, j_3=1/2}$ sequentially calculating the knot invariant.

\paragraph{$R$-matrix for different spin}

For calculation the knot invariant with it is critically important to define the matrix 
$(R^{j_1,j_2,j_3})^{n_1,n_2,n_3}_{m_1,m_2,m_3}$ for different spins $j_1,j_2,j_3$. As an example, take $R^{j_1=1/2, j_2=1, j_3=1/2}$, where the eigenvalues of \eqref{Knot18}, depending on the spectral parameter, are
\begin{eqnarray}
\lambda_0(u)=\sinh(\mu+u),\nonumber\\
\lambda_\frac{1}{2}(u)=\sinh(\frac{3\mu}{2}+u),\nonumber\\
\lambda_1(u)=\sinh(\mu-u),\nonumber\\
\lambda_\frac{3}{2}(u)=\sinh(\frac{3\mu}{2}-u),
\end{eqnarray}
\begin{eqnarray}
\lambda_0(u)\stackrel{u\longrightarrow\infty}{\rightarrow}(\frac{e^{u+\mu}}{2})(\frac{e^{u+2\mu}}{2})=\frac{e^{2u}}{2^{2}}e^{3\mu}=\left\{(\frac{e^{2u}}{4})q^{\frac{3}{2}}\right\}(1),\\
\lambda_\frac{1}{2}(u)\stackrel{u\longrightarrow\infty}{\rightarrow}(\frac{e^{u+\mu}}{2})(\frac{-e^{u-2\mu}}{2})=\frac{e^{2u}}{2^{2}}(-e^{-\mu})=\left\{(\frac{e^{2u}}{4})q^{\frac{3}{2}}\right\}(-q^{-1}),\\
\lambda_1(u)\stackrel{u\longrightarrow\infty}{\rightarrow}(\frac{-e^{u-\mu}}{2})(\frac{e^{u+2\mu}}{2})=\frac{e^{2u}}{2^{2}}(-e^{\mu})=\left\{(\frac{e^{2u}}{4})q^{\frac{3}{2}}\right\}(-q^{2}),\\
\lambda_\frac{3}{2}(u)\stackrel{u\longrightarrow\infty}{\rightarrow}(\frac{-e^{u-  \mu}}{2})(\frac{-e^{u-2\mu}}{2})=\frac{e^{2u}}{2^{2}}(-e^{-\mu})=\left\{(\frac{e^{2u}}{4})q^{\frac{3}{2}}\right\}(-q^{-1}),
\end{eqnarray}
The eigenvalues of the braids for three strands carrying spins $\frac{1}{2},1,\frac{1}{2}$, their representation in $SU(2)$ of the Chern-Simons theory have the form
\begin{eqnarray}
\lambda^{B}_{1}\sim(-)^{l}q^{\frac{-l(l+1)}{2}},\\
\lambda^{B}_{0}\sim 1,\ \ \lambda^{B}_{1/2}\sim-q^{-1},\ \ \lambda^{B}_{l}\sim-q^{2},\ \ \lambda\stackrel{B}{\frac{3}{2}}\sim q^{-1}.
\end{eqnarray}
Here $\lambda _l (u)$ are proportional to $\lambda^{B}_{l}$ knots $u\rightarrow (u)$ 
\begin{eqnarray}
M=m_1+m_2=n_1+n_2
\end{eqnarray}
and the general form of the $R$-matrix will be
{\begin{eqnarray}
(R^{\frac{1}{2},1,\frac{1}{2}})^{n_1,n_2,n_3}_{m_1,m_2,m_3}(u)=\nonumber\\
\resizebox{.9\textwidth}{!}
{$\begin{pmatrix}
 m_1,m_2/n_1,n_2 &\vline&\frac{1}{2},1,\frac{1}{2}&\frac{1}{2},1,-\frac{1}{2}&\frac{1}{2},-1,-\frac{1}{2}&\frac{1}{2},0,\frac{1}{2}&\frac{1}{2},0,-\frac{1}{2}&-\frac{1}{2},0,\frac{1}{2}&-\frac{1}{2},0,-\frac{1}{2}&-\frac{1}{2},1,\frac{1}{2}&-\frac{1}{2},-1,\frac{1}{2}&-\frac{1}{2},-1,-\frac{1}{2}\\
\hline\frac{1}{2},1,\frac{1}{2}&\vline&
\chi_1(u)&0&0&0&0&0&0&0&0&0\\ \frac{1}{2},1,-\frac{1}{2}&\vline&  
0&\chi_2(u)&0&\chi^{'}_{3}(u)&0&0&0&0&0&0&\\ \frac{1}{2},-1,-\frac{1}{2}&\vline&
0&0&\chi_{2}(u)&0&\chi^{'}_5(u)&0&\chi^{'}_6(u)&0&0&0\\ \frac{1}{2},0,\frac{1}{2}&\vline&
0&\chi_3(u)&0&\chi_4(u)&0&0&0&\chi^{'}_6(u)&0&0\\   \frac{1}{2},0,-\frac{1}{2}&\vline&
0&0&\chi_5(u)&0&\chi^{'}_{4}(u)&0&0&0&0&0\\-\frac{1}{2},0,\frac{1}{2}&\vline& 
0&0&0&0&0&\chi^{'}_{4}(u)&0&\chi^{'}_{5}(u)&0&0\\ -\frac{1}{2},0,-\frac{1}{2}&\vline& 
0&0&\chi_{6}(u)&0&0&0&\chi_{4}(u)&0&\chi^{'}_{3}(u)&0\\  -\frac{1}{2},1,\frac{1}{2}&\vline&
0&0&0&\chi_{6}(u)&0&\chi_5(u)&0&\chi_{2}(u)&0&0\\  -\frac{1}{2},-1,\frac{1}{2}&\vline&
0&0&0&0&0&0&\chi_3(u)&0&\chi_{2}(u)&0\\  -\frac{1}{2},-1,-\frac{1}{2}&\vline&
0&0&0&0&0&0&0&0&0&\chi_{1}(u)
\end{pmatrix}$},
\end{eqnarray}}

where
\begin{eqnarray}
\chi_1=2\sinh(\frac{3\mu}{2}-u)+\sinh(\mu-u), \chi_2=\cosh 2\mu-\sinh u,\nonumber\\ \chi_3=(\sinh 2\mu \sinh\mu)^{\frac{1}{2}}e^{-u}, \chi_4=2\sinh(\frac{\mu}{2}-u)+\sinh(\mu-u), \nonumber\\ \chi_5=-\sinh 2\mu(\sinh 3\mu)^{\frac{1}{2}}\sinh(\frac{3\mu}{2}+u)-\sinh u, \nonumber\\ \chi_6=\sinh 2\mu(\sinh\mu)^{\frac{1}{2}}e^{u}-(\sinh 3\mu)^{\frac{1}{2}}\sinh(\frac{3\mu}{2}), \nonumber\\ \chi^{'}_3=(\sinh 2\mu\sinh\mu)^{\frac{1}{2}}e^{u}, \chi^{'}_5=\sinh 2\mu(\sinh 3\mu)^{\frac{1}{2}}\sinh(\frac{3\mu}{2}+u)+\sinh u, \nonumber\\ \chi^{'}_6=\sinh 2\mu(\sinh\mu)^{\frac{1}{2}}e^{-u}+(\sinh 3\mu)^{\frac{1}{2}}\sinh(\frac{3\mu}{2}).
\end{eqnarray}
Replacing $u\rightarrow\infty$ and $q=e^{2\mu}$ for $j_1=1/2, j_2=1, j_3=1/2$, we have

\begin{eqnarray}
\lim_{u\rightarrow \infty}\frac{{(R^{\frac{1}{2},1,\frac{1}{2}})}^{n_1,n_2,n_3}_{m_1,m_2,m_3}(u)}{(R^{\frac{1}{2},1,\frac{1}{2}})^{\frac{1}{2},1,\frac{1}{2}}_{\frac{1}{2},1,\frac{1}{2}}(u)}=\nonumber\\ 
\resizebox{.9\textwidth}{!}{$\begin{pmatrix}
 m_1,m_2/ n_1,n_2 &\vline&\frac{1}{2},1,\frac{1}{2}&\frac{1}{2},1,-\frac{1}{2}&\frac{1}{2},-1,-\frac{1}{2}&\frac{1}{2},0,\frac{1}{2}&\frac{1}{2},0,-\frac{1}{2}&-\frac{1}{2},0,\frac{1}{2}&-\frac{1}{2},0,-\frac{1}{2}&-\frac{1}{2},1,\frac{1}{2}&-\frac{1}{2},-1,\frac{1}{2}&-\frac{1}{2},-1,-\frac{1}{2}\\
\hline\frac{1}{2},1,\frac{1}{2}&\vline&
1&0&0&0&0&0&0&0&0&0\\ 
\frac{1}{2},1,-\frac{1}{2}&\vline&  
0&\frac{q^\frac{1}{4}-q^{-\frac{1}{4}}}{q^\frac{1}{2}-q^{-\frac{1}{2}}}&0&-\frac{q^\frac{3}{4}-q^{-\frac{3}{4}}}{q^\frac{1}{2}-q^{-\frac{1}{2}}}&0&0&0&0&0&0&\\ \frac{1}{2},-1,-\frac{1}{2}&\vline&
0&0&\frac{q^\frac{1}{4}-q^{-\frac{1}{4}}}{q^\frac{1}{2}-q^{-\frac{1}{2}}}&0&q^\frac{1}{2}-q^{-\frac{1}{2}}&0&-q-q^{-1}&0&0&0\\ \frac{1}{2},0,\frac{1}{2}&\vline&
0&\frac{q^\frac{3}{4}-q^{-\frac{3}{4}}}{q^\frac{1}{2}-q^{-\frac{1}{2}}}&0&-q^\frac{1}{4}-q^{-\frac{1}{4}}&0&0&0&-q-q^{-1}&0&0\\ \frac{1}{2},0,-\frac{1}{2}&\vline& 
0&0&-q^{\frac{1}{2}}-q^{-\frac{1}{2}}&0&q^{\frac{1}{4}}-q^{-\frac{1}{4}}&0&0&0&0&0\\-\frac{1}{2},0,\frac{1}{2}&\vline& 
0&0&0&0&0&q^{\frac{1}{4}}-q^{-\frac{1}{4}}&0&q^{\frac{1}{2}}-q^{-\frac{1}{2}}&0&0\\ -\frac{1}{2},0,-\frac{1}{2}&\vline& 
0&0&q-q^{-1}&0&0&0&-q^{\frac{1}{4}}-q^{-\frac{1}{4}}&0&-\frac{q^\frac{3}{4}-q^{-\frac{3}{4}}}{q^\frac{1}{2}-q^{-\frac{1}{2}}}&0\\  -\frac{1}{2},1,\frac{1}{2}&\vline&
0&0&0&q-q^{-1}&0&-q^{\frac{1}{2}}-q^{-\frac{1}{2}}&0&\frac{q^\frac{1}{4}-q^{-\frac{1}{4}}}{q^\frac{1}{2}-q^{-\frac{1}{2}}}&0&0\\  -\frac{1}{2},-1,\frac{1}{2}&\vline&
0&0&0&0&0&0&\frac{q^\frac{3}{4}-q^{-\frac{3}{4}}}{q^\frac{1}{2}-q^{-\frac{1}{2}}}&0&\frac{q^\frac{1}{4}-q^{-\frac{1}{4}}}{q^\frac{1}{2}-q^{-\frac{1}{2}}}&0\\  -\frac{1}{2},-1,-\frac{1}{2}&\vline&
0&0&0&0&0&0&0&0&0&1
\end{pmatrix}$}.
\end{eqnarray}

To obtain the braid generators $b(j_1={\frac{1}{2}}, j_2=1, j_3={\frac{1}{2}})$, choose the corresponding permutation $\hat{P}^{j_1={\frac{1}{2}}, j_2=1, j_3={\frac{1}{2}}}$ so that the sequence of spins indicated in the row and column above in $R^{j_1={\frac{1}{2}}, j_2=1, j_3={\frac{1}{2}}}$-the matrix remains unchanged. For $j_1={\frac{1}{2}}, j_2=1, j_3={\frac{1}{2}}$, the $\hat{P}^{j_1={\frac{1}{2}}, j_2=1, j_3={\frac{1}{2}}}$ maybe 
\begin{eqnarray}
\resizebox{.8\textwidth}{!}{$\hat{P}^{j_1={\frac{1}{2}}, j_2=1, j_3={\frac{1}{2}}}=
\begin{pmatrix}
1& 0& 0& 0&  0& 0& 0& 0& 0& 0&\\
0& 0& 1& 0&  0& 0& 0& 0& 0& 0&\\
0&  0& 0& 0&  1& 0& 0& 0& 0& 0&\\
0&  0& 0& 0&  0& 0& 1& 0& 0& 0&\\
0&  0& 0& 0&  0& 0& 0& 0& 1& 0&\\
0& 1& 0& 0&  0& 0& 0& 0& 0& 0&\\
0& 0& 0& 1&  0& 0& 0& 0& 0& 0&\\
0&  0& 0& 0&  0& 1& 0& 0& 0& 0&\\
0&  0& 0& 0&  0& 0& 0& 1& 0& 0&\\
0&  0& 0& 0&  0& 0& 0& 0& 0& 1&\\
\end{pmatrix}$}.
\end{eqnarray}
We apply this permutation matrix in the construction of the braid generator
\begin{eqnarray}
(\hat{R}^{j_1,j_2,j_3})_{m_1,m_2,m_3}^{n_1,n_2,n_3}=(\hat{P}^{j_1,j_2,j_3})^{{m^{'}_{1},m^{'}_{2},m^{'}_{3}}}_{{m_{1},m_{2},m_{3}}}\lim_{u\rightarrow \infty}\frac{{(R^{\frac{1}{2},1,\frac{1}{2}})}^{n_1,n_2,n_3}_{m_1,m_2,m_3}(u)}{(R^{\frac{1}{2},1,\frac{1}{2}})^{\frac{1}{2},1,\frac{1}{2}}_{\frac{1}{2},1,\frac{1}{2}}(u)},
\end{eqnarray}
as a result of which we obtain the explicit form $\hat{R}^{j_1={\frac{1}{2}}, j_2=1, j_3={\frac{1}{2}}}$-matrix for the three-partite vertex model
\begin{eqnarray}(\hat{R}^{\frac{1}{2},1,\frac{1}{2}})_{m_1,m_2,m_3}^{n_1,n_2,n_3}=\nonumber\\
\resizebox{.9\textwidth}{!}{$
\begin{pmatrix}
1& 0& 0& 0&  0& 0& 0& 0& 0& 0&\\
0& 0& \frac{-\frac{1}{q^{\frac{1}{4}}}+q^{\frac{1}{4}}}{-\frac{1}{\sqrt{q}+\sqrt{q}}}& 0& -\frac{1}{q}-q& 0& 0& 0& 0& 0&\\
0&  0& -\frac{1}{\sqrt{q}}-\sqrt{q}& 0&  -\frac{1}{q^{\frac{1}{4}}}+q^{\frac{1}{4}}& 0& 0& 0& 0& 0&\\
0&  0& -\frac{1}{q}+q& 0& 0&  0&  -\frac{1}{q^{\frac{1}{4}}}+q^{\frac{1}{4}}& 0& -\frac{-\frac{1}{q^{\frac{3}{4}}}+q^{\frac{
3}{4}}}{-\frac{1}{\sqrt{q}+\sqrt{q}}}& 0&\\
0&  0& 0& 0&  0& 0& \frac{-\frac{1}{q^{\frac{3}{4}}}+q^{\frac{
3}{4}}}{-\frac{1}{\sqrt{q}+\sqrt{q}}}& 0& \frac{-\frac{1}{q^{\frac{1}{4}}}+q^{\frac{1}{4}}}{-\frac{1}{\sqrt{q}+\sqrt{q}}}& 0&\\
0& \frac{-\frac{1}{q^{\frac{1}{4}}}+q^{\frac{1}{4}}}{-\frac{1}{\sqrt{q}+\sqrt{q}}}& 0& -\frac{-\frac{1}{q^{\frac{3}{4}}}+q^{\frac{3}{4}}}{-\frac{1}{\sqrt{q}+\sqrt{q}}}&  0& 0& 0& 0& 0& 0&\\
0& \frac{-\frac{1}{q^{\frac{3}{4}}}+q^{\frac{3}{4}}}{-\frac{1}{\sqrt{q}+\sqrt{q}}}& 0& -\frac{1}{q^{\frac{1}{4}}}-q^{\frac{1}{4}}&  0& 0& 0& -\frac{1}{q}-q& 0& 0&\\
0&  0& 0& 0&  0& -\frac{1}{q^{\frac{1}{4}}}+q^{\frac{1}{4}}& 0& -\frac{1}{\sqrt{q}}+\sqrt{q}& 0& 0&\\
0&  0& 0& -\frac{1}{q}+q&  0& -\frac{1}{\sqrt{q}}-\sqrt{q}& 0& \frac{-\frac{1}{q^{\frac{1}{4}}}+q^{\frac{1}{4}}}{-\frac{1}{\sqrt{q}+\sqrt{q}}}& 0& 0&\\
0&  0& 0& 0&  0& 0& 0& 0& 0& 1&\\
\end{pmatrix}$}.
\end{eqnarray}
Similar construction $\hat{R}^{\frac{1}{2},1}$ for $\frac{1}{2},1,\frac{1}{2}$ is the transposed matrix $\hat{R}^{1,\frac{1}{2}}$. The explicit form of the identity matrix can be written as
\begin{eqnarray}
\hat{R}^{j_1,j_2}[\hat{R}^{j_1,j_2}]^{-1}=\hat{R}^{j_1,j_2}\left[[\hat{R}^{j_2,j_1}]^\intercal\right]^{-1}=\mathbb{I},\\
\hat{R}^{j_1,j_3}[\hat{R}^{j_1,j_3}]^{-1}=\hat{R}^{j_1,j_3}\left[[\hat{R}^{j_3,j_1}]^\intercal\right]^{-1}=\mathbb{I},\\
\hat{R}^{j_2,j_3}[\hat{R}^{j_2,j_3}]^{-1}=\hat{R}^{j_2,j_3}\left[[\hat{R}^{j_3,j_2}]^\intercal\right]^{-1}=\mathbb{I}.
\end{eqnarray}

		Thus, we have the opportunity to represent the matrix forms $b_i[j_1,j_2,j_3]$ groupoids that satisfy the constructions of the braid generators
\begin{eqnarray}
b_i[{j_1,j_2}][b_i^{j_1,j_2}]([b_i^{j_2,j_1}])^{-1}\mathbb{I},\\
b_i[{j_1,j_3}][b_i^{j_1,j_3}]([b_i^{j_3,j_1}])^{-1}\mathbb{I},\\
b_i[{j_2,j_3}][b_i^{j_2,j_3}]([b_i^{j_3,j_2}])^{-1}\mathbb{I},
\end{eqnarray}
in this form
\begin{eqnarray}
b_i[{j_1,j_2}]=\underbrace{\mathbb{I}\times\mathbb{I}\times\ldots\mathbb{I}\times}_{i-1}(\hat{R}^{j_1,j_2})^{-1}\times\mathbb{I}_{i+2},\\
b_i[{j_1,j_3}]=\underbrace{\mathbb{I}\times\mathbb{I}\times\ldots\mathbb{I}\times}_{i-1}(\hat{R}^{j_3,j_1})^{-1}\times\mathbb{I}_{i+2},\\
b_i[{j_2,j_3}]=\underbrace{\mathbb{I}\times\mathbb{I}\times\ldots\mathbb{I}\times}_{i-1}(\hat{R}^{j_3,j_2})^{-1}\times\mathbb{I}_{i+2}.
\end{eqnarray}
As a result, using the matrix representation, for any word of the braid $A$, the closure of which will give multicomponent links at the knot. Similarly to the formula for the knot invariant \eqref{Knot4}, we present the following formula for the multi-colored knot invariant, in which the components of the knot have different spins.
Invariant of multi-colored knot $\tilde{\alpha}_{j_{1},j_{2},j_{3}}(A)$ (up to the total deformation coefficient $q^{1/2}$) for a six-component link, where the components of the knot have different spins, obtained by closing any $r$-word of the braid CollapseBraid $A$ is given by formula
\begin{eqnarray}
\alpha_{j_{1},j_{2},j_{3}}[A(CB)]=q^{\frac{1}{2}C}\tilde{\alpha}_{j_{1},j_{2},j_{3}}(A)=q^{\frac{1}{2}C}\prod^{n}_{i=1}(\tau_{j_{i}}\overline{\tau_{j_{i}}})^{-\frac{l_{i}}{2}}Tr\left\{HA\right\},
\end{eqnarray}
where the first factor derives from $q$ the coefficient of deformation with an integer $C$, which depends on the spins, the bending and  number between constituent knots of links. The $l_{i}$'s is the number of strands in the braid generator, that is, when the spin $j_{i}$ is in the $r$-strand braid $A$, where $\sum_{i=1}^{n}l_{i}=r$. In addition, the matrix representation of $H$ depends on the order of such repeating spins occurring in a braid of $r$-strands. This article describes a $3$-strand braid obtained from knot $5_2$, with $j_1$ braiding on the first strand, $j_2\neq j_1$ on the second, and $j_3$ on the third strand, which implies
\begin{eqnarray}
H=h_{j_{1}}\otimes h_{j_{2}}\otimes h_{j_{3}}.
\end{eqnarray}
Coming back to the matrix operations for the braid word $A$, we use the formula \eqref{Knot14}, following the sequence of closure of each link in the braid, where each strand has the same spins mentioned above. From the above definition $h_{j_{i}}$'s \eqref{Knot5}, $\tau_{j_{i}}$'s and $\overline\tau_{j_{i}}$'s \eqref{Knot6}, as discussed in section $2$. In the next section, we will calculate in detail the invariant of the multi-colored link $5_2$ knot.

\paragraph{Multi-colored knot invariants}

In this section of the work, we will find the invariant of a knot with links with spins $j_1=j_3=\frac{1}{2}$ and $j_2=1$, for this we first need to write a braid word of $n$-strands that tracks the spins. 
For the CollapseBraid [BR] Knot $5_2$ knot, consisting of simple links obtained by closing a three-strand braid and the matrix representation looks like this
\begin{eqnarray}
A([BR]Knot 5_2)=b_2^{-1}(\frac{1}{2},\frac{1}{2})b_2(1,\frac{1}{2})b_3^{-1}(\frac{1}{2},1)b_2^{-1}(\frac{1}{2},1)b_3^{-1}(\frac{1}{2},1)b_2^{-1}(1,\frac{1}{2}),\nonumber\\
A([BR]Knot 5_2)=b_2^{-1}(\frac{1}{2},\frac{1}{2})b_2(1,\frac{1}{2})b_3^{-1}(\frac{1}{2},1)b_2^{-1}(\frac{1}{2},1)b_3^{-1}(\frac{1}{2},1)b_2^{-1}(1,\frac{1}{2})
\end{eqnarray}
and $H=h_{\frac{1}{2}}\otimes h_{1}\otimes h_{\frac{1}{2}}$ will give us
\begin{eqnarray}
\alpha_{\frac{1}{2},1,\frac{1}{2}}[A(5_2)]={\frac{\kappa\delta{\eta}^{\frac{3}{2}}}{\psi}},
\end{eqnarray}
 where 

$\kappa=\sqrt{q}$, 

$\eta=(\frac{(1+q+q^{2}+q^{3}+2q^{4}+q^{5}+q^{6}+q^{9}+q^{11}+q^{13})({1+q+q^{2}+q^{3}+2q^{4}+q^{5}+q^{6}+q^{9}+q^{11}+q^{13}})}{q^{13}})$,

$\psi=((1+\sqrt{q})^{5}q^{\frac{5}{2}}(1+q+q^{2}+q^{3}+2q^{4}+q^{5}+q^{6}+q^{9}+q^{11}+q^{13}))$,

$\delta=1-q^\frac{1}{4}+3\sqrt{q}-4q^{\frac{3}{4}}+2q-2q^{\frac{5}{4}}-7q^{\frac{3}{2}}+15q^{\frac{7}{4}}-13q^{2}+40q^{\frac{9}{4}}+2q^{\frac{5}{2}}+52q^{\frac{11}{4}}+\nonumber\\+32q^{3}+44q^{\frac{13}{4}}+76q^{\frac{7}{2}}+40q^{\frac{15}{4}}+150q^4+55q^{\frac{17}{4}}+238q^{\frac{9}{2}}+48q^{\frac{19}{4}}+289q^5-15q^{\frac{21}{4}}+\nonumber\\+276q^{\frac{11}{2}}-104q^{\frac{23}{4}}+192q^6-142q^{\frac{25}{4}}+27q^{\frac{13}{2}}-75q^{\frac{27}{4}}-174q^7+88q^{\frac{29}{4}}-328q^{\frac{15}{2}}+\nonumber\\+312q^{\frac{31}{4}}-335q^8+466q^{\frac{33}{4}}-241q^{\frac{17}{2}}+422q^{\frac{35}{4}}-167q^9+223q^{\frac{37}{4}}-80q^{\frac{19}{2}}+37q^{\frac{39}{4}}+\nonumber\\+22q^{10}-40q^{\frac{41}{4}}+18q^{\frac{21}{2}}-34q^{\frac{43}{4}}-119q^{11}+30q^{\frac{45}{4}}-253q^{\frac{23}{2}}+151q^{\frac{47}{4}}-304q^{12}+\nonumber\\+264q^{\frac{49}{4}}-292q^{\frac{25}{2}}+262q^{\frac{51}{4}}-231q^13+137q^{\frac{53}{4}}-133q^{\frac{27}{2}}+2q^{\frac{55}{4}}-4q^{14}-30q^{\frac{57}{4}}+\nonumber\\+105q^{\frac{29}{2}}+46q^{\frac{59}{4}}+130q^{15}+136q^{\frac{61}{4}}+95q^{\frac{31}{2}}+148q^{\frac{63}{4}}+47q^{16}+84q^{\frac{65}{4}}+9q^{\frac{32}{2}}+\nonumber\\+9q^{\frac{67}{4}-14q^{17}}-21q^{\frac{69}{4}}-21q^{\frac{35}{2}}-5q^{\frac{71}{4}}-15q^{18}+10q^{\frac{73}{4}}-6q^{\frac{37}{2}}+8q^{\frac{75}{4}}-q^{19}+2q^{\frac{77}{4}}$,\\
which is consistent with the multi-colored Jones polynomial calculated on the basis of the $SU(2)$ Chern-Simons theory up to a full factor.

\section{Conclusion}

In this work, we have obtained a completely new $R$-matrix of the three-partite vertex model for the knot $5_2$ with multi-colored links from the representations of the braid group, produced by the Vogel algorithm. Here, adjacent knot edges bear different spins. This invariant is proportional to the multi-colored polynomial Jones. In the article the representation of the group $SO(N)$ instead of $SU(2)$ was used, which has spins $j_1,j_2,j_3$ states on the edges intersecting the vertex. The process of finding the knot invariant outlined in the paper should be generalizable for the group $SO(N)$ and generate, in turn, new vertex models and their invariants. In the future, one can find such types of vertex models and solutions of their knot invariants with different links. The found invariant can be compared with the HOMFLY-PT polynomial indicated in the literature.

\section*{Acknowledgments}

This study was funded by the Science Committee of the Ministry of Education and Science of the Republic of Kazakhstan AP08955524.

\bibliography{mybibfile}

\end{document}